\documentclass{ieeeaccess}
\usepackage{amsmath,amsfonts,amssymb,amsthm,bbm,graphicx,enumerate,times,mathtools,physics,marvosym,wasysym}
\usepackage{hyperref}
\hypersetup{colorlinks=true,citecolor=black,linkcolor=black,urlcolor=blue}
\usepackage{float}
\usepackage[ruled,longend]{algorithm2e}
\usepackage{etoolbox}
\usepackage[noabbrev,capitalise]{cleveref}

\creflabelformat{equation}{#2#1#3}
\SetKw{KwGenerate}{Generate:}
\SetKw{KwAssign}{Assign:}
\SetKw{KwCalculate}{Calculate:}

\begin{document}
\history{Accepted for Publication in IEEE Transactions on Quantum Engineering}
\doi{10.1109/TQE.2020.DOI}

\title{Topological-Graph Dependencies and Scaling Properties of a Heuristic Qubit-Assignment Algorithm}
\author{\uppercase{Matthew Steinberg}\authorrefmark{1,2,3}, \uppercase{Sebastian Feld}\authorrefmark{1,2}, \uppercase{Carmen G. Almudever}\authorrefmark{1,4}, \uppercase{Michael Marthaler}\authorrefmark{3}, and \uppercase{Jan-Michael Reiner}\authorrefmark{3}}
\address[1]{QuTech, Delft University of Technology, Delft, the Netherlands}
\address[2]{Quantum and Computer Engineering Department, Delft University of Technology, Delft, the Netherlands}
\address[3]{HQS Quantum Simulations GmbH, Haid-und-Neu-Straße 7, D-76131 Karlsruhe, Germany}
\address[4]{Computer Engineering Department, Technical University of Valencia, Valencia, Spain}
\corresp{Corresponding author: Matthew Steinberg (email: \url{matt.steinberg3@gmail.com}).}

\date{\today}

\begin{abstract}
The qubit-mapping problem aims to assign and route qubits of a quantum circuit onto a NISQ device in an optimized fashion, with respect to some cost function. Finding an optimal solution to this problem is known to scale exponentially in computational complexity; as such, it is imperative to investigate scalable qubit-mapping solutions for NISQ computation. In this work, a noise-aware heuristic qubit-assignment algorithm (which assigns initial placements for qubits in a quantum algorithm to qubits on a NISQ device, but does not route qubits during the quantum algorithm's execution) is presented and compared against the optimal \textit{brute-force} solution, as well as a trivial qubit assignment, with the aim to quantify the performance of our heuristic qubit-assignment algorithm. We find that for small, connected-graph algorithms, our heuristic-assignment algorithm faithfully lies in between the effective upper and lower bounds given by the brute-force and trivial qubit-assignment algorithms. Additionally, we find that the topological-graph properties of quantum algorithms with over six qubits play an important role in our heuristic qubit-assignment algorithm's performance on NISQ devices. Finally, we investigate the scaling properties of our heuristic algorithm for quantum processors with up to 100 qubits; here, the algorithm was found to be scalable for quantum-algorithms which admit path-like graphs. Our findings show that as the size of the quantum processor in our simulation grows, so do the benefits from utilizing the heuristic qubit-assignment algorithm, under particular constraints for our heuristic algorithm. This work thus characterizes the performance of a heuristic qubit-assignment algorithm with respect to the topological-graph and scaling properties of a quantum algorithm which one may wish to run on a given NISQ device.
\end{abstract}


\maketitle

\section{Introduction}

Quantum computation may still be in its infancy, but new advances have allowed for the first experimental demonstrations of quantum computing in recent years \cite{google,blattion1,64IBM,chinesephotonic,2020demonstrationHoneywell}. Quantum computers themselves promise to aid in solving classically-intractable problems for such fields as quantum chemistry \cite{quantum_chem1,quantum_chem2,benjamin1,Bromley_2020}, quantum machine learning \cite{biamonte1,QMLreview}, quantum field theory \cite{jordan2019quantum,2012preskill,funcke2021quantum,brennen2021nearly}, and quantum cryptography \cite{quantumcrypto,PiraQCrypto}, among others. However, such promise comes with a catch: protecting the quantum states in a quantum computer from deleterious noise channels has proven to be a most difficult task, preventing scalability and implementations of most quantum algorithms \cite{PreskillNISQ}. Indeed, current prototypes of \textit{quantum processing units} (or QPUs) available from Rigetti, Honeywell, IBM, Google, Intel \cite{sete-zeng-rigetti,2020demonstrationHoneywell,64IBM,google,intel1,intel2} and others are considered still too resource-constrained to be able to demonstrate full fault-tolerant quantum computation \cite{QEC1,2020egan-brown,Linke_2017}.   

In the \textit{Noisy Intermediate-Scale Quantum} (or NISQ) era, quantum computers are still rapidly evolving but still present several limitations: Firstly, quantum computing as a field has not yet settled on a particular physical realization for quantum hardware \cite{fullstack}; leading candidate implementations include those constructed from superconducting qubits \cite{superconducting1,blueprintsuper,superFT,chinesesuper,couplingsuper,sete-zeng-rigetti}, trapped-ion qubits \cite{ions,ions2,ions3,ions4}, as well as other proposals \cite{silicon,photons,Bourassa_2021,bartolucci2021fusionbased,chamberland2020buildingconcatenated}. Secondly, many devices exhibit fixed and finite connectivity constraints between neighboring qubits (a notable exception to this is trapped-ion technology, in which one can \textit{in principle} produce "all-to-all" connectivity \cite{ions}). Thirdly, noise considerations have severely hampered developments in quantum devices \cite{knillNoise}. As such, efficient methods for executing quantum algorithms on first-generation quantum hardware require special attention. 

In light of these difficulties, the goal of efficiently delegating these finite resources in a QPU for usage with near-term quantum algorithms is both exigent and pervasive. Quantum algorithms described as quantum circuits have to be adapted to specific hardware constraints in order to be executed. Many different approaches exist for realizing this aim, which range widely from algorithm compilation \cite{fullstack} and neural-network based approaches \cite{ML1,ML2}, to more theoretically-motivated methods such as quantum gate-synthesis techniques adapted for quantum hardware \cite{gate-synthesis,gate-synthesis2,wille1,wille2}. 

Recent work has centered on the \textit{qubit-mapping problem}, which aims to answer the following question: Given a resource-constrained quantum device and a prospective quantum algorithm, what is the optimal strategy for assigning (known as the \textit{assignment of qubits}) and coordinating movements (known as \textit{quantum-state routing}) of qubits along the lattice of the QPU, while making guarantees on the algorithm's fidelity? Many methods of solving this problem have been studied, wherein the most-promising approaches so far have explicitly taken into account error information from the quantum device (e.g. two-qubit and single-qubit gate fidelities) \cite{aliroquantum,hardware-aware,lao-mapping1,ML1,ML2,noisy1,temporal1,qubitproblem1}. In spite of this progress, much remains to be done in order to understand the general characteristics of such qubit-mapping techniques. Due to the computational hardness of finding optimized solutions for the qubit-mapping problem, one commonly resorts to heuristic algorithms \cite{noisy1,hardware-aware}. Therefore, assessing the capabilities of heuristic qubit-mapping algorithms is essential for addressing scalability concerns in NISQ hardware. As the assignment of qubits is a fundamental part of the qubit-mapping process \cite{noisy1,siraichi:hal-01655951}, this work will focus on this first step using a heuristic qubit-assignment algorithm (HQAA). For a thorough review of the qubit-mapping problem, we refer the reader to \cite{bandic,carmina_overview}.

Currently, heuristic qubit-assignment algorithms lack a systematic comparative basis by which their performance can be assessed; this includes the notion of providing upper and lower bounds for the performance of a quantum algorithm executed on a NISQ device, evaluating quantum circuits whose entangling two-qubit gates produce interaction graphs with different topological graph-theoretic structures, and understanding whether or not a given heuristic qubit-assignment algorithm will scale well as both the quantum algorithm and the QPU size are increased.

The purpose of this manuscript is to answer the following questions: $i)$ how does a HQAA compare relative to an optimal brute-force and a trivial approach to the qubit-assignment problem, $ii)$ do the topological features of a quantum algorithm influence the measured success rate for a noise-aware HQAA, and $iii)$ What behavior can one expect from a noise-aware HQAA as the size of the QPU is scaled up? To this end, we develop an HQAA which is similar to the noise-aware one introduced in \cite{noisy1}, but improve upon this work by incorporating notions of \textit{graph centrality} \cite{centrality1,centrality2}. We run tests with several realistic benchmarks in order to provide approximate upper and lower bounds to the success rate of the HQAA using implementations of brute-force and trivial assignment algorithms (BFAA and TAA, respectively). We systematically analyze an HQAA with respect to topologically-inequivalent graph representations of quantum circuits, and assign them to an $n \times n$ QPU lattice which we keep constant throughout most of the study. An analysis of topologically-inequivalent quantum-circuit structures is provided, and it is shown that the topological characteristics of two quantum algorithms (with identical numbers of gates) indeed play a role in the average success rate measured for our HQAA, when quantum algorithms consisting of more than six qubits are examined. Finally, we investigate the scaling properties of the HQAA for quantum algorithms whose gate structure gives rise to path-like graph representations (which commonly appear in simulations of fermionic quantum systems \cite{quantum_chem1,benjamin1,quantum_chem2}); we find that our HQAA consistently exhibits higher success rates than a TAA, when considering path-like quantum algorithms on larger quantum-device architectures, as long less than $75\%$ of the quantum processor is filled. We additionally find that the benefits of utilizing our HQAA increase with the size of the QPU.

The structure of this paper is as follows: \cref{background-qmapping-section} provides a background to the qubit-mapping problem and a walk-through of a basic example. \cref{description_algs-section} provides details on the structure of each of the algorithms employed; additionally, we show how to calculate the success rate for our simulations. We separate our results in  \cref{results-intro-section} into several parts: We describe the benchmarks utilized in our analysis in \cref{real-benchmarks-section}; next, we discuss the results obtained from incorporating non-nearest-neighbor two-qubit gates in the benchmarks we tested (\cref{nonlinear-section}), using and comparing a \textit{breadth versus depth} analysis of two-qubit gate additions in order to generate topologically-inequivalent quantum-algorithmic graphs; and finally, in \cref{large-linear-section} we examine the outcomes of our simulations with quantum-algorithmic path-like graph structures, scaled onto QPU connectivity graphs with dimensions $n \times n$ qubits, where $n > 3$. In \cref{discussion-section}, we conclude our work and discuss possible future directions. 

\section{Background}\label{background-qmapping-section}

Most quantum circuits that are devised theoretically do not take into account the actual physical hardware constraints of a given quantum device. In order to accommodate NISQ hardware, quantum-programming frameworks such as Qiskit \cite{Qiskit} include supports which allow developers to write algorithms without explicitly factoring in hardware limitations. As such, quantum compilers must perform several steps in order to prepare the quantum algorithm for actual execution on a device. Broadly speaking, these steps are: $1)$ to decompose the quantum gates into \textit{elementary gates}; $2)$ to assign the qubits of the quantum circuit to the physical qubits of the quantum processor, and inserting SWAP operations (known as \textit{routing}) into the algorithm in order to satisfy the connectivity constraints of a given quantum device; $3)$ and finally, to optimize the resultant quantum circuit, with an aim to minimize quantities such as the execution time and gate count, among other cost functions. The qubit-mapping problem consists of the second step of quantum compilation and will be the main focus of this article, with an emphasis on the \textit{assignment of the qubits} of a quantum algorithm to the physical qubits of a quantum processor. 

In the context of the qubit-mapping problem, we consider two objects: an \textit{interaction graph} (IG), which is a graph representation of the quantum circuit that we would like to execute (wherein vertices represent qubits, and its weights represent the number of single-qubit gates invoked; conversely, the edges correspond to two-qubit gate interaction, with weights designating the quantity of such interactions), and the \textit{coupling graph} (CG), which is a graph representation of the QPU's geometric connectivity (wherein vertices represent the physical qubits used in the device, edges represent the two-qubit gate interactions which are possible, and weights on vertices and edges represent the single and two-qubit gate errors). For the purposes of this manuscript, we define a \textbf{graph} to be an ordered pair $G(V,E)$ of two sets known as the \textit{vertex set} $V$ and \textit{edge set} $E$. The sets of vertices $V$ and edges $E$ are interpreted \textit{qubits} and \textit{two-qubit gate interactions}, respectively; this relationship is displayed in \cref{fig1}. Two graphs are termed \textbf{topologically equivalent} if there is a map $f: X \mapsto Y$ between two graphs $X$ and $Y$ such that the following conditions are upheld \cite{gross1987topological}:

\begin{enumerate}
\item The map $f$ is bijective, i.e. $f$ maps from all edges to all edges and from all vertices to all vertices;
\item $f$ is continuous, i.e. $f$ is an isomorphism from $X$ to $Y$, allowing for the graph operations of \textit{smoothing out} and \textit{subdivision} of edges; and
\item The inverse function $f^{-1}$ is continuous. 
\end{enumerate}

These criteria do not form the centerpiece of the present work; rather, the notion of topologically-equivalent graphs will be useful for understanding the rest of the paper.

Most realistic QPU layouts are accompanied with \textit{noise-calibration statistics} which are added to the graph; these data usually include two-qubit gate error rates, single-qubit error rates, execution times (gate length), relaxation energies, and decoherence characteristic times $T1$ and $T2$ \cite{hardware-aware}. The goal is to match the geometric connectivity of the IG to that of the CG as closely as possible (in effect defining a \textit{graph isomorphism} in the case of an exact match \cite{gross1987topological}) while taking into account the noise-calibration statistics of the quantum device as well. In the present work, we do not directly utilize the noise-calibration statistics from a real quantum computer; instead, we have analyzed the statistics from several of the IBM quantum computers \cite{64IBM,Murali_2020software,noisy1}, and assume random errors on the same order of magnitude (which are typically $\sim 10^{-3}$ for single-qubit errors, and $\sim 10^{-2}$ for two-qubit gate errors and measurement errors \cite{noisy1}), which are then randomly assigned to nodes and edges on the QPU CG. These errors are utilized in a cost function for evaluating the success rate of the quantum algorithm, using our HQAA. The procedure for this analysis is discussed in detail in \cref{description_algs-section} . Additionally, several proposals show that qubits can be given \textit{initial assignments}, which later may be modified in timestep fashion as the execution of the algorithm progresses. As the basis for this work considers only the \textit{initial assignment} for the qubit-mapping problem, we refer the reader to \cite{temporal1,qubitproblem1,lao-mapping1,Tan_2020-medina-talk} for work involving \textit{routing/time-scheduling techniques}.

\begin{figure*}
\centering
\hskip-0.5cm
\includegraphics[width=\textwidth]{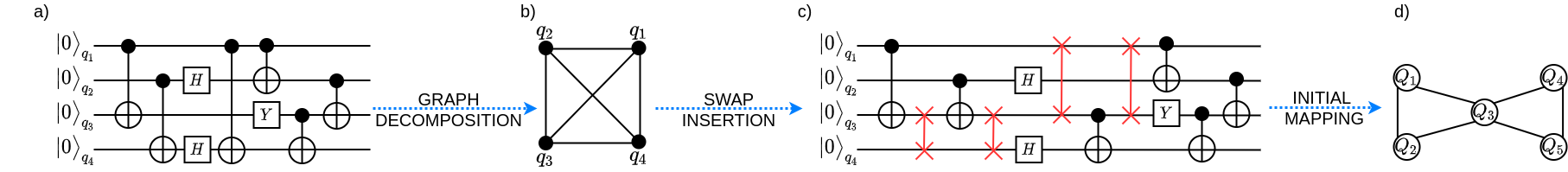}
\caption{a) depicts an example quantum circuit. As shown in b), this circuit is decomposed into a graph-theoretic version of the algorithm (also known as an \textit{interaction graph} (IG)) which illustrates the interaction via two-qubit gates of qubits in the original algorithm; edges represent two-qubit gates, while vertices represent qubits which are acted upon. Weights are added to vertices and edges in order to account for more gate invocations. In c) the geometric connectivity of the IG is analyzed and SWAP gates are added for any interaction-graph edges which cannot be exactly assigned to the QPU coupling graph (CG); for example, if we define an assignment $\{q_{1}\mapsto Q_{1}, q_{2}\mapsto Q_{2},q_{3} \mapsto Q_{3} \}$, then the interaction-graph edge between $q_{1}$ and $q_{4}$ cannot be explicitly assigned, since $Q_{1}$ and $Q_{4}$ do not share an edge connection in the QPU CG shown in d). The modified quantum circuit is then assigned to the QPU in d); in this way, the appropriate vertices, in accordance with some metric to be defined, are assigned to the graph-theoretic object representing the quantum device (referred to as the CG of the QPU). In many types of qubit-mapping algorithms, qubits can then be arranged and mapped \textit{temporally}, as well as \textit{spatially} during the \textit{routing process} \cite{hardware-aware,lao-mapping1,qubitproblem1}; for this reason, the final arrow between c) and d) carries the designation \textit{initial assignment}.}
\label{fig2}
\end{figure*}

In order to illustrate how the qubit-assignment problem can be treated, consider the quantum algorithm in \cref{fig2}a. Before assigning qubits from the quantum algorithm to the physical QPU, the corresponding circuit is itself decomposed into a graph-theoretic form which we defined as the \textit{interaction graph}; such a decomposition is needed in order to properly assign virtual qubits to a relevant portion of the QPU lattice such that the geometric constraints of the circuit are respected. As shown in \cref{fig2}b, the resulting IG is the \textit{complete graph} $K_{4}$, and cannot be exactly embedded into the QPU CG shown in \cref{fig2}d; equivalently, one may say that no structure-preserving map (i.e. a \textit{graph isomorphism}) $f$ exists from the IG to the CG \cite{gross1987topological}. In order to correctly assign this algorithm, one may add a SWAP gate operation to the qubits $q_{1}$ and $q_{3}$, and then perform the required two-qubit gate between $q_{3}$ and $q_{4}$, as depicted in the modified algorithm of \cref{fig2}c; other SWAP gates are added as well in \cref{fig2}. The SWAP gate itself degrades the final-state fidelity of the algorithm, in accordance with the commensurate two-qubit gate error rates. Due to the disadvantages of utilizing SWAP gates, many qubit-mapping algorithms explicitly attempt to minimize the amount of SWAP gates employed \cite{bandic,lao-mapping1}. However, we did not explicitly design our algorithm with such a notion in mind, even though necessary SWAP gates are considered.

Solutions to both the qubit-assignment and qubit-mapping problems can be separated into two broad categories: $1)$ \textit{optimal} (or brute-force) optimization and $2)$ heuristic optimization algorithms \cite{bandic}. As described before, current studies of qubit-mapping and assignment focus on minimizing the number of SWAP gates \cite{lao-mapping1,bandic,Tan_2020-medina-talk,palerwille,carmina_overview}. However, our particular solution to the qubit-assignment problem is not the main focus of the present manuscript; rather we concentrate on the evaluation of topologically-inequivalent IGs for an HQAA, building and expanding on the work pioneered by \cite{noisy1}, with an aim towards understanding the topological-graph properties of IGs and how they influence the performance of an HQAA, while keeping the QPU CG \textit{effectively constant}.

\Figure[t!](topskip=0pt, botskip=0pt, midskip=0pt)[scale=0.35]{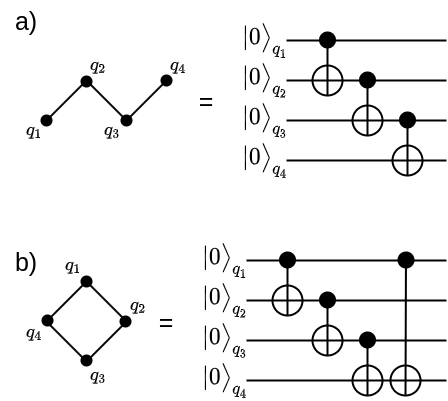}
{In this work, we introduce the notions of IGs which are \textit{path-like} and \textit{cycle-like}. a) displays a nearest-neighbor quantum circuit, with a path-like IG; b) shows that adding a cyclic edge to the IG is equivalent to the addition of an extra two-qubit gate, which is not nearest-neighbor in the corresponding quantum circuit.\label{fig1}}

For the present work, we introduce the concepts of IGs which are \textit{path-like} and \textit{cycle-like} in the context of the qubit-assignment problem in order to efficiently assess the topological-graph structure of a quantum algorithm. Two examples of this idea can be seen in \cref{fig1}. In \cref{fig1}a, we take a four-vertex IG with three edges (known as a \textit{Hamiltonian path} $P_{4}$) to be equivalent to a four-qubit quantum circuit exhibiting nearest-neighbor two-qubit gates. As is shown in \cref{fig1}b, if an extra edge is added, the IG becomes a \textit{cycle graph} $C_{4}$, and corresponds to the addition of an extra two-qubit gate between the first and last qubits in the quantum circuit. In graph theory, the two objects in \cref{fig1} are well known; the problem of identifying suitable \textit{Hamiltonian-path} and \textit{cycle-graph} solutions in a simple undirected graph is related to the \textit{traveling salesman problem} and is known to be NP-Complete \cite{wilson1,combinatorial,2021deterministicnearestneighbor}; such computational complexity necessitates our present study. Later, we shall further seek to specify interaction-graph structure by introducing edges to a path-like IG either in a \textit{breadth-first} or \textit{depth-first} approach, resulting in several pairing of IGs which have the same number of edges, but are \textit{topologically-inequivalent}, cycle-like IGs\footnote[1]{One may also consider \textit{tree-like} IGs; however, for the purposes of this manuscript, we will concentrate on path-like and cycle-like IGs, and reserve this discussion for further study.}. 

In this article, we shall focus primarily on the two differing cases described above: quantum algorithms whose IGs admit $1)$ path-like and $2)$ cycle-like forms (such IGs are briefly discussed in \cref{real-benchmarks-section}). The emphasis on path-like IGs is justified for two main reasons. Firstly, in fields such as quantum chemistry, the simulation of fermionic quantum systems can be carried out by encoding the qubits via a \textit{Jordan-Wigner transformation} \cite{quantum_chem1,benjamin1,quantum_chem2}; such an encoding scheme can give rise to circuits known as \textit{linear SWAP networks} \cite{quantum_chem3,quantum_chem4}, exhibiting path-like IGs. Secondly, a quantum algorithm with a path-like interacting-graph representation is, in a sense, hardware agnostic; such algorithms exhibit only nearest-neighbor two-qubit gate invocations, and are thus adaptable to any architecture. Designing a qubit-assignment algorithm for the goal of considering such quantum algorithms is therefore paramount. Conversely, our precise motivation for investigating \textit{graph cyclicity} as it applies to the qubit-assignment problem is not motivated necessarily with respect to realistic implementations, but rather as a method by which to probe the limits of when our HQAA effectively fails to find an adequate solution. In this way, we set forth a method by which to diagnose and analyze the effectiveness of a general HQAA with respect to a diverse plethora of interaction-graph topological structures, whilst keeping the coupling-graph connectivity constant; moreover, we utilize a similar framework to investigate the scaling properties of our HQAA, in particular as the size of the QPU increases. In the following section, we will detail our approach for the qubit-assignment algorithm, specifically made with the previously-mentioned goals in mind. 
 
\section{Description of the Qubit-Assignment Algorithms}\label{description_algs-section}

All of the qubit-assignment algorithms utilized in this work generally function in the following manner: firstly, an $n \times n$ square lattice (representing the geometric connectivity of the QPU) is initialized, along with single-qubit and two-qubit error rates, as well as measurements, as a NetworkX object \cite{networkx} which will serve as an approximation for the QPU device's CG. Next, a quantum algorithm written in cQASM \cite{cQASM} is parsed into a NetworkX object as well. The qubit-assignment algorithm is then called, a final-assignment solution is assigned, and the assignment is evaluated using a cost function that is described at the end of this section (we will refer to this cost function as the \textit{metric}). The IGs used in this article for QPU simulations were assigned to CGs which consist of $3 \times 3$ QPU lattice grids in the simulations from \cref{nonlinear-section}; for the large path-like simulations of \cref{large-linear-section}, we assign to $n \times n$ grid lattice QPUs qubits, where $n > 3$. All of the code for this project is freely accessible on Github\footnote[2]{\textbf{\url{https://github.com/mattsteinberg13/heuristic-qubit-mapping-algorithm}}}.

For all of the algorithms described in this section, we will refer to an IG, comprised of some set of vertices $V$ and some set of edges $E$, as $\tilde{I}(V,E)$; additionally, we shall refer to a QPU CG with vertex set $V'$ and edge set $E'$ as $\tilde{Q}(V',E')$. For the algorithms that we designed, several assumptions were made: 

\begin{enumerate}
\item If $\Delta(V) \geq \Delta(V')$ (where $\Delta(V)$ represents the \textit{maximal vertex degree} \cite{wilson1}), then an assignment solution exists for $\tilde{I}(V,E)$, which may or may not require SWAP gates.
\item We assume that $|V| \leq |V'|$ (where $|V|$, $|V'|$ represent the total number of vertices in the interaction and CGs, respectively), i.e. that the number of vertices in the IG is smaller than or equal to the number of vertices in the CG. If $|V| > |V'|$, then the assignment process aborts, and an error message is displayed. An example of such an error would concern the assignment of an $n$-qubit quantum algorithm to an $(n-1)$-qubit CG.
\end{enumerate}

\subsection{The HQAA \& Traffic Coefficient}\label{heuristic-description-section}

Our HQAA is greedy in nature \cite{greedy}, and pseudocode for the algorithm is described in \cref{alg2}. The HQAA functions generally as follows. The IG $\tilde{I}(V,E)$ and the CG $\tilde{Q}(V',E')$ are initialized as NetworkX objects after being parsed from an input QASM file. Next, the set of traffic coefficients $V_{i, \text{tc}}$ and the maximal traffic coefficient $V_{i, \text{tc}}^{\text{max}}$ are calculated and stored in lists, which can be interpreted as the overall percentage of gate invocations for a given qubit. These mathematical objects will be explained in more detail in the next two paragraphs. Subsequently, the \textit{infimum} vertex (in the present context we refer to the 'infimum-vertex qubit' as the maximal-degree qubit with the minimal two-qubit error-rate edge on the CG, which we denote in \cref{alg2} as $\inf_{\Delta(V'),\mathcal{E}(E')}$, where $\Delta(V')$ and $\mathcal{E}(E')$ represent the maximal-degree and the minimal two-qubit error-rate edge for the CG, respectively) of the CG is selected as the first candidate coupling-graph qubit to be assigned to. Afterwards, our algorithm defines an initial assignment from the interaction-graph qubit with the maximal traffic coefficient to the infimum coupling-graph qubit; we cycle through both ordered sets of interaction-graph and coupling-graph qubits, defining neighboring interaction-graph qubits in a nearest-neighbor style on the QPU CG. When all nearest neighbors for a given coupling-graph vertex have been defined already, our algorithm uses Dijkstra's shortest-path algorithm in order to find the physically closest coupling-graph qubit (here we take the 'shortest path' to mean the edge(s) that constitute the overall lowest edge error rate); as such, whenever a nearest-neighbor qubit is not able to be located, our policy is to add SWAP gates, in accordance with the shortest path that the Dijkstra's algorithm finds. This process is continued until all of the interaction-graph qubits have been matched to a corresponding coupling-graph qubit. We will provide more detail in the rest of the section.

\begin{algorithm}
\KwIn{$\tilde{I}(V,E)$, $\tilde{Q}(V',E')$}
\tcp{Get traffic coefficients (see Eq. \ref{eq:eq1})}
$V_{i, \text{tc}} \gets f_{i}(N_{s,i},N_{d,i})$ \\ 
\tcp{Get infimum-vertex CG-qubit} 
$V'_{\text{inf}} \gets \inf_{\Delta(V'),\mathcal{E}(E')} V'$ \\
\tcp{Get IG-qubit with maximal tc}
$V_{i, \text{tc}}^{\text{max}} \gets \max V_{i, \text{tc}}$ \\
\tcp{Perform initial assignment}
$V_{i, \text{tc}}^{\text{max}} \mapsto V'_{\text{inf}}$ \\ 
\tcp{Remember as last-assigned CG qubit}
$V'_{\text{last-assigned}} \gets V'_{\text{inf}}$ \\
\tcp{Assign remaining IG-qubits} 
\For{$v \in V_{i,tc}\backslash V_{i,tc}^{max}$}{
\tcp{Identify next candidate CG-qubit}
\eIf{NN\_CG\_qubit\_exists$(V'_{\text{last-assigned}})$}{
\tcp{find NN CG-qubit}
$V'_{N} \gets$ get\_NN\_CG\_qubit$(V'_{\text{last-assigned}})$ \\}(){
\tcp{If all NN $V'_{N}$ assigned, find NCN}
$V'_{N} \gets$ get\_NCN\_CG\_qubit$(V'_{\text{last-assigned}})$\\}{}
\tcp{Perform consecutive assignment}
$v \mapsto V'_{N}$ \\
\tcp{Remember as last-assigned CG qubit}
$V'_{\text{last-assigned}} \gets V'_N$ \\
}
\caption{Pseudocode for the HQAA described in \cref{heuristic-description-section}. Here, we use the abbreviations \textit{nearest-neighbor} (NN) and \textit{next-closet neighbor} (NCN) in the "if" statement below.}\label{alg2}
\end{algorithm}

The HQAA presented here is necessarily similar to the simple heuristic assignment strategy used in \cite{noisy1}; however, our HQAA is novel in the sense that we utilize the traffic coefficients as a fitness function for the interaction-graph qubits. As mentioned earlier, the notion of traffic coefficients is related to the notion of \textit{vertex centrality} in graph theory \cite{centrality1,centrality2}. Additionally, \cite{noisy1} evaluates and compares several heuristic qubit-assignment algorithms to SMTP-based algorithms. We stress here that our aim is not to make significant improvements to state-of-the-art qubit-mapping strategies, but rather to directly quantify and qualify topological-graph dependencies between an interaction graph and our HQAA, as well as investigating the scaling properties of our HQAA. We suspect that our findings will have wider applicability to most qubit-mapping algorithms, although we will comment on this in \cref{discussion-section}.

The maximal traffic coefficient is calculated as follows. First, for the i$^{\text{th}}$ interaction-graph qubit, we sum the total number of single- and two-qubit gate invocations; the \textit{frequency} of an interaction-graph qubit is subsequently labeled $f_{i}$, as shown in \cref{eq:eq1}. Here, we have weighted two-qubit gate invocations ($N_{d,i}$) with an extra linear multiplier of $2$ in order to weigh interaction-graph qubits which exhibit a large percentage of the two-qubit gates utilized in a given quantum algorithm more heavily than single-qubit gates. In this way, we take into account for \textit{both} the nontriviality of the single-qubit gate invocations ($N_{s,i}$) and the higher error rates of two-qubit gates, which are typically at least one order of magnitude worse than for single-qubit gates \cite{Murali_2020software,noisy1}. After a bit of algebra (\cref{eq:eq2,eq:eq3}), one sees that the frequency of the $i^{\text{th}}$ interaction-graph is re-written as a \textit{traffic coefficient} $v \in V_{i,\text{tc}}$; these traffic coefficients are summed and normalized, such that $cv$ provides a "percentage-wise" overview of the total interactions for the $i^{\text{th}}$ interaction-graph qubit in an algorithm. We then take the \textit{maximal traffic coefficient} $V_{i, \text{tc}}^{\text{max}}$ as corresponding to the first interaction-graph qubit to be assigned, as shown in \cref{eq:eq4}.

\begin{align}
f_{i} = N_{s,i} + 2N_{d,i} \label{eq:eq1},\\
1 - \frac{1}{f_{i}} = V_{i, \text{tc}} \label{eq:eq2},\\
c \cdot \sum_{i} V_{i,\text{tc}} = 1 \label{eq:eq3},\\
\max V_{i, \text{tc}} = V_{i, \text{tc}}^{\text{max}} \label{eq:eq4},
\end{align}

As a brief aside, one may ask why the single-qubit error rates are factored in at all, given the large difference in magnitude between the two-qubit and single-qubit error rates. The reason for this is the following: consider a quantum circuit, where the ratio of single-qubit gates to double-qubit gates is much higher than $1$. Given such a scenario, we expect that the error rates of the single-qubit gates non-trivially factor into the determination of which virtual qubit should be first allocated.

The maximal traffic coefficient $V_{i, \text{tc}}^{\text{max}}$ represents the percentage of interactions for the "most active" qubit in the algorithm, and is used as a way to ascertain which interaction-graph qubits must be prioritized for the best-connected, lowest error-rated portions of the QPU CG via our HQAA that is described in \cref{heuristic-description-section}. It is entirely possible that more than one interaction-graph qubit may have the largest traffic coefficient; in this case, we simply iterate through the set of qubits with maximal traffic coefficient $V_{i, \text{tc}}^{\text{max}}$, and then treat the rest of the qubits in the assignment process.

In order to assigned the rest of the IG, a variant of Dijkstra's algorithm \cite{dijkstra} is utilized, which takes into account the error rates of two- and single-qubit gates (represented on the CG as edges and vertices with assigned error rates) in order to find the "shortest path" (in this case the term \textit{shortest path} refers to the particular sequence of gates which lead to the lowest error rate, which is the shortest path, since weights are placed on the edges as two-qubit error rates) to the next available qubit in the IG; once the next candidate interaction-graph qubit is designated, the algorithm surveys the interaction-graph qubits that have already been assigned. Finally, the HQAA assigns the new candidate to the QPU CG, as closely as possible (such that the least amount of errors is generated) to the originally-assigned interaction-graph qubit. This process continues until the entire algorithm has been assigned to the closest-possible qubits on the IG.

\subsection{The Brute-Force and Trivial Qubit-Assignment Algorithms}

The BFAA utilized in this work functions as follows: first, the lattice QPU is initialized, and the error rates for all quantities are defined; next, the BFAA generates a list of all possible permutations for a quantum-algorithm mapping solution. Each permutation is assigned, and is evaluated using the metric described in \cref{success-rate-metric-explain}. The preceding permutation's metric value is compared to the current iteration, and the permutation with the highest success-rate metric value is kept, while the inferior one is discarded. This process continues until the best permutation is found.

\Figure[t!](topskip=0pt, botskip=0pt, midskip=0pt)[width=3.5cm]{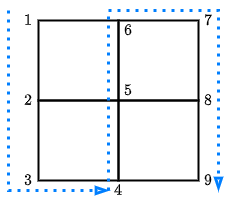}
{The numbering scheme employed by our TAA follows a "snake pattern" across the CG.\label{fig3}}

The TAA functions by sequentially assigning interaction-graph qubits to correspondingly-numbered coupling-graph qubits; a quantum algorithm with qubits $q_{1}\dots q_{r}$, where $r \leq n^{2}$ ($n^{2}$ for an $n \times n$ QPU), will be assigned to a CG of a QPU with qubits $Q_{1}\dots Q_{s}$, where $s \geq r$, by assigning interaction-/coupling-graph qubit pairs as $\{q_{1} \mapsto Q_{1},q_{2} \mapsto Q_{2},\dots, q_{r} \mapsto Q_{r}\}$. The success-rate metric is then subsequently evaluated, in order to compare with the other two assignment strategies. In all likelihood, one can imagine many different numbering schemes that will give rise to differing evaluations of the success rate; as such, we fixed our numbering scheme for the QPU CG in order to follow a snake pattern along the processor's topology, as shown \cref{fig3}. The numbering scheme used for the IGs can be considered to be arbitrary, and follows from the numbering of the qubits in its respective qubit register. As such, our TAA iteratively assigns interaction-graph qubits to a CG while following a snake pattern.

\subsection{Evaluation of the Success-Rate Metric}\label{success-rate-metric-explain}

The cost function used to quantify the performance of all algorithms in this work is described below. The purpose of this metric is to approximate the fidelity of the final quantum state after the quantum circuit is assigned and executed, with all gates invoked, in a computationally efficient manner. Other metrics exist \cite{lao-mapping1,noisy1,qubitproblem1,temporal1}; however, we limit our attention here to a metric that is based on success-rate measures. 

The single-qubit gate, two-qubit gate, and SWAP-gate product metrics are calculated as shown below in \cref{eq:single-metric,eq:double-qubit-metric,eq:swap-product,eq:product_metric_total}. These product metrics, as explained above, relate specifically to the CGs that we utilize in this work. 

\begin{align}
\sigma_{s} = \prod_{i}^{n'<n}(1-\xi_{s,i})^{N_{s,i}} \label{eq:single-metric},\\
\sigma_{d} = \prod_{i}^{\delta_{d}}(1-\xi_{d,i})^{N_{d,i}} \label{eq:double-qubit-metric}, \\
\sigma^{\text{SW}} = \prod_{j}^{\delta^{\text{SW}}}\prod_{i}^{l}(1-\xi^{\text{SW}}_{i})_{j}^{(2N^{\text{SW}}_{i})} \label{eq:swap-product}, \\
\sigma_{\text{total}} = \sigma_{s} \cdot \sigma_{d} \cdot \sigma^{\text{SW}} \label{eq:product_metric_total},
\end{align}

where on the first line, $\sigma_{s},n',n,\xi_{s,i},N_{s,i}$ are: the total single-qubit gate metric value; the total number of qubits in the IG; the total number of qubits in the CG; the single-qubit gate error rate for each qubit; and the number of single-qubit gate invocations per qubit, respectively. On the second line, $\sigma_{d},\delta_{d},\xi_{d,i},N_{d,i}$ are: the total two-qubit gate metric value; the total number of edges on the NISQ device; the two-qubit gate error rate per edge; and the number of two-qubit gate invocations per edge, respectively. On the final line, $\sigma^{\text{SW}},\delta^{\text{SW}},l,\xi^{\text{SW}}_{i},2N^{\text{SW}}_{i}$ represent: the total SWAP-gate metric value; the total number of separate edges that need SWAP gates; the total number of edges which physically separate the SWAP-corrected pair of coupling-graph qubits; the two-qubit gate error rate per edge; and the number of two-qubit gate invocations, for which the error rates overall are squared (this takes into account the cost of moving the qubit both to and from the closest unoccupied region, using the SWAP gate), respectively. Finally, $\sigma_{\text{total}}$ represents the total metric calculated from the product of all other success rates, as mentioned above. For simplicity, we do not explicitly take into account how SWAP gates may be invoked on real QPU devices \cite{wille1,wille2}.

Since the HQAA itself functions by allocating interaction-graph qubits to coupling-graph qubits based on a measure of the success-rate, one may ask how useful it is to utilize a metric based on the same measures. The reason for this can be seen as follows: when considering the optimal solution for any qubit-assignment algorithm, there are a variety of different cost functions that one may utilize. However, \textit{any assignment} that is based on an exact calculation of the fidelity of the final quantum state after running it on a physical QPU will give the best indication of performance for an assignment solution \cite{nishio-errors}. Since direct computations of the fidelity are computationally resource-intensive and scale exponentially with the size of the QPU \cite{PreskillNISQ,nielsen-chuang-qit}, we opt to utilize a cost function that may be regarded as related to the calculation of the fidelity (i.e.\ a success rate measurement, based on the error rates as presented). In this way, we attempt to employ a cost function that is as relevant as possible, without the computational demands incurred by large-scale simulations. 

In agreement with \cite{noisy1}, a precise formulation of the noise model is not warranted in this work. One main reason for this discrepancy is due to the inherent nature of the cost function used, which takes into account the experimentally-determined noise-calibration statistics, not the specific details of the quantum channels acting upon the system. Further details are not necessary, as such inclusions will impact the scalability of the qubit-assignment algorithm.

In the next section, we will discuss in full the results obtained from studying several realistic interaction-graph benchmarks, benchmarks whose IGs exhibit high degrees of cycle-like edges, and benchmarks which increase in size and sequentially occupy more and more of a given QPU coupling-graph's qubits. 

\section{Results}\label{results-intro-section}

The results are organized as follows: \cref{real-benchmarks-section} details the results obtained from realistic benchmark IGs, assigned to a $3 \times 3$ CG. \cref{nonlinear-section} discusses the results from assigning quantum benchmarks with IGs that exhibit increasing amounts of non-nearest neighbor two-qubit gate combinations. The results from scaling the size of path-like IGs onto ever-increasing QPU CGs are described in \cref{large-linear-section}. All benchmarks were tested on a Dell Latitude $7400$ laptop with a $1.9$ GHz x $4$ Intel i7-8665U quadcore processor and $8.0$ GB of RAM. Each benchmark was assigned using our simulation 100, 100, and 1000 times in \cref{real-benchmarks-section,nonlinear-section,large-linear-section}, respectively; success rates were averaged over all trials. Simulations of Sequences I and II (as shown in \cref{fig6}) took approximately 60 hours of continuous runtime. 

\subsection{Realistic Interaction-Graph Benchmarks}\label{real-benchmarks-section}

\begin{figure}[htp]
\centering
\includegraphics[width=\columnwidth]{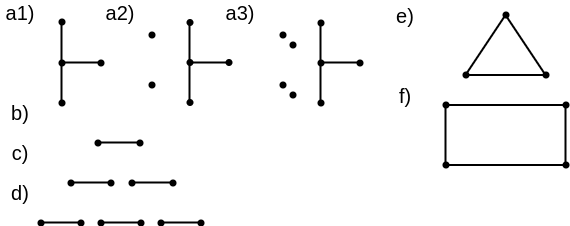}
\caption{IGs of several realistic benchmarks which were tested in \cref{real-benchmarks-section}; the benchmarks themselves were taken from \cite{noisy1}. a1)-a3) show the BV4, BV6, and BV8 benchmarks, respectively; b) represents the QFT and HS2 benchmark algorithms; the IGs in c) and d) were used for the HS4 and HS6 algorithms; e) depicts the Fredkin, Or, Peres, and Toffoli algorithms; and f) displays the Adder benchmark.}
\label{fig4}
\end{figure}

\begin{figure}[htp]
\centering
\includegraphics[width=\columnwidth]{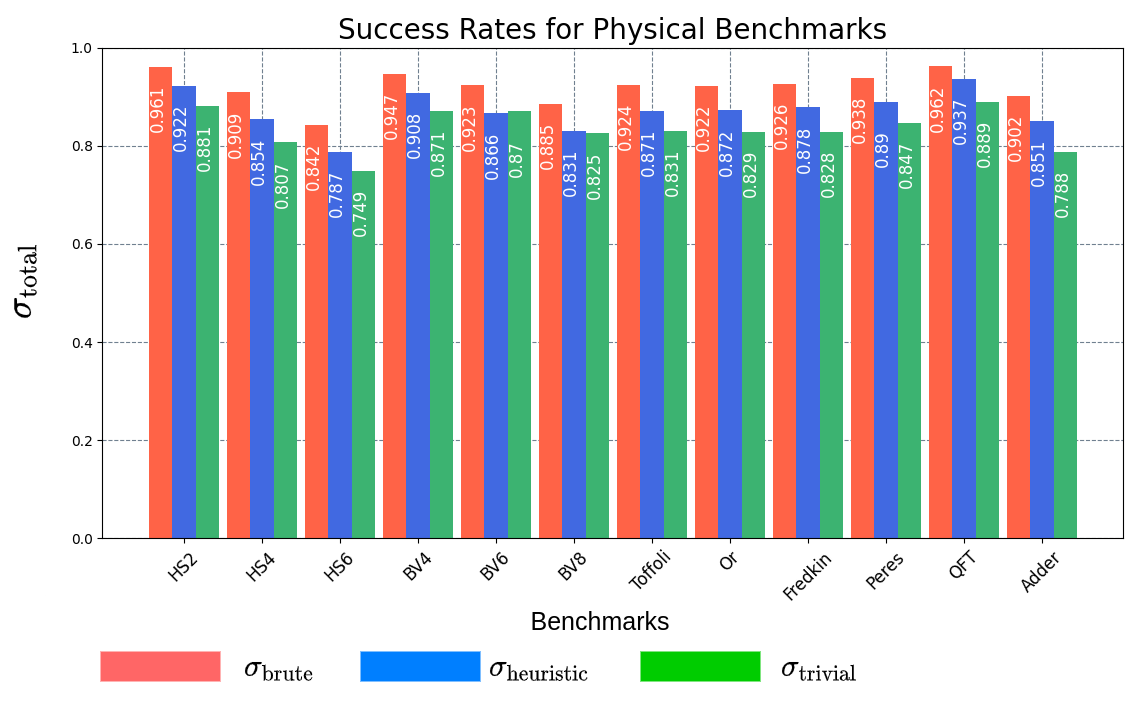}
\caption{Success rate for the benchmarks in \cite{noisy1}. The BFAA results are shown in red, followed by blue and green bars, which represent the HQAA and TAA results, respectively.}
\label{fig5}
\end{figure}

The benchmarks from \cite{noisy1} were utilized in this section. The corresponding IGs for each of the benchmarks listed are shown in \cref{fig4}. All of the benchmarks were tested on a $3 \times 3$ lattice CG, and the results are reported in \cref{fig5}. In this section, we did not explicitly take into account whether the IGs are path-like or cycle-like; path-like and cycle-like interaction-graph benchmarks will be investigated and compared in detail in \cref{nonlinear-section,large-linear-section}.

The results obtained from the BFAA, HQAA, and TAA with respect to the benchmarks detailed in \cref{fig4} are shown in \cref{fig5}. The algorithms themselves are grouped: the first three groupings of results pertain to the \textit{hidden-shift algorithmic} benchmarks; the second three groupings relate the results obtained for the \textit{Bernstein-Vazirani} benchmarks; as for the next grouping, the results belonging to the \textit{Toffoli, Or, Fredkin}, and \textit{Peres} benchmarks exhibit triangular IGs; and finally, the \textit{Quantum Fourier Transform} (QFT) and \textit{Adder} benchmarks admit IGs as shown in \cref{fig4}b and \cref{fig4}f, respectively. Red bars denote the results from the BFAA; blue and green bars denote the results from the commensurate HQAA and TAA, respectively. The y-axis of \cref{fig5} displays the calculated success rates. 

As is evidenced in the graphs, the BFAA provides an effective upper bound for the performance of the HQAA (we use the term \textit{effective upper bound} with the view that better solutions may exist if one utilizes \textit{time-scheduling techniques}, as mentioned in \cref{background-qmapping-section}); additionally, the TAA allows for an interpretation as an effective lower bound, with the HQAA pivoting between both of these effective bounds. The HQAA outperforms the TAA in virtually all cases except for those involving disjointness in the interaction-graph quantum algorithms, as indicated by the success-rate values of \cref{fig4} a2) and \cref{fig4} a3).

One may na{\"i}vely conclude from the results obtained that the HQAA used in this work performs generally well for all types of interaction-graph topologies, both cycle-like and path-like\footnote[3]{The only exception to this rule can be seen in the \textit{disjoint} IGs that were utilized for the HS4 and HS6 algorithms. Such a complication is hardly unexpected, as our HQAA was intended for connected-IGs only. One can in principle consider additions in order address quantum algorithms that prepare product states; however, we reserve the discussion on disjoint IGs for future work.}. However, one may expect that topological-graph effects play a limited role in such small-scale quantum algorithms. Therefore, our results from this section serve to motivate a more systematic investigation of the topological-graph effects of an IG on the performance of our HQAA. We will show in the next section that topological-graph structure plays a role in the performance of our HQAA for IGs larger than six virtual qubits, and thus were not apparent in the results of this section. 

\subsection{Topologically-Inequivalent Interaction-Graph Benchmarks}\label{nonlinear-section}

\begin{figure*}
\centering
\includegraphics[width=\textwidth]{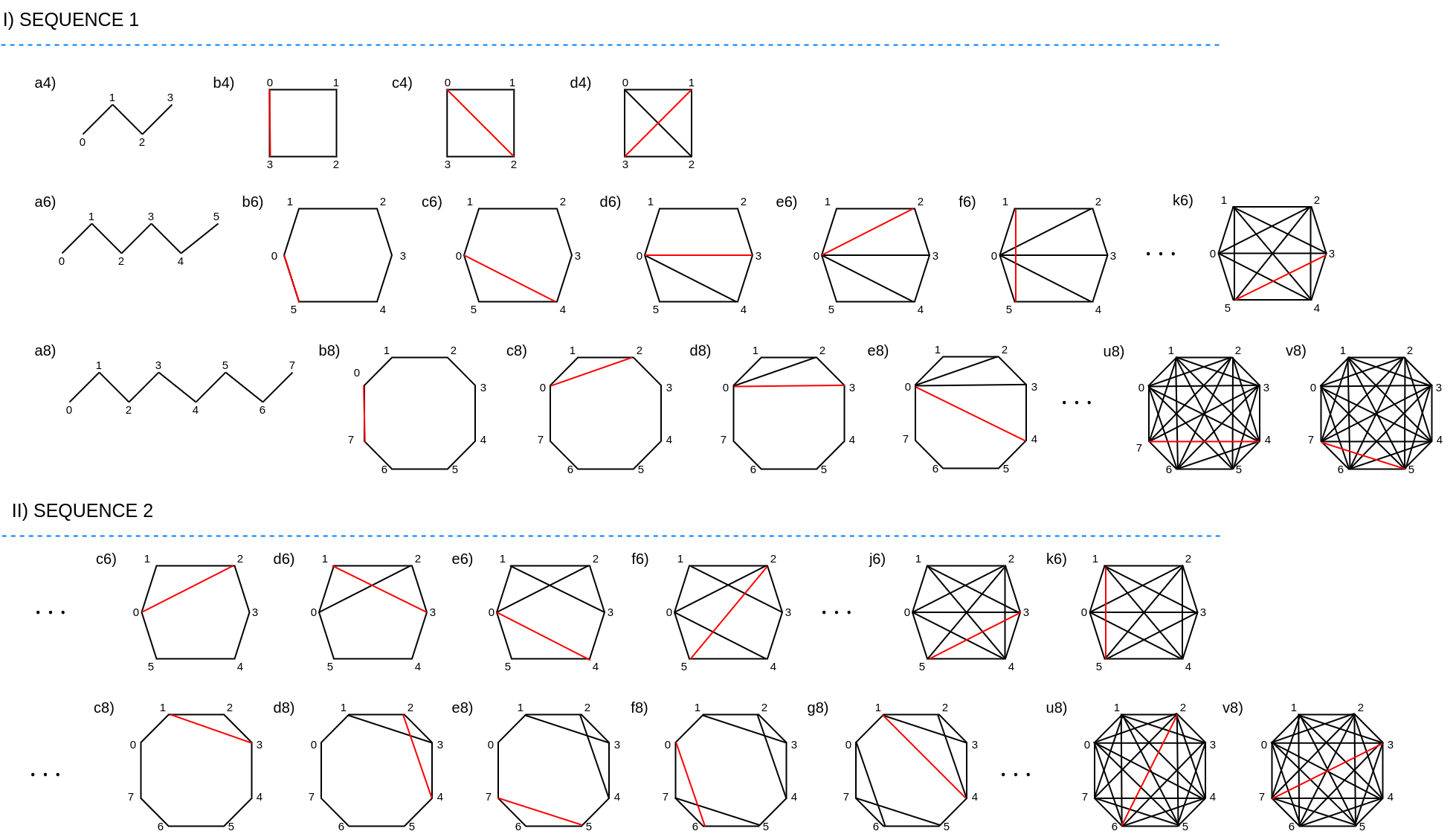}
\caption{IGs for the first and second benchmark sequences. Red edges depict newly-added cycle-like edges to the IGs that were tested with our HQAA. In the IGs from Sequence I, cycle-like edges are appended such that every vertex degree is maximized before subsequently appending cycle-like edges to consecutively higher-numbered vertices; this process continues until a $K_{s}$ graph is attained for Sequence I (We refer to this technique as a \textit{depth-first} edge-assignment procedure). As for Sequence II, a \textit{breadth-first} edge-assignment procedure is utilized, as cycle-like edges are appended such that every vertex exhibits approximately the same degree until a $K_{s}$ graph is reached.}
\label{fig6}
\end{figure*}

The results of this section detail a comparison between the HQAA, BFAA, and TAA. These two additional qubit-assignment algorithms provide \textit{effective} upper and lower bounds on the performance of the HQAA, just as in the previous section. $4$-, $6$-, and $8$-qubit IGs were used as benchmarks to be assigned onto a $3 \times 3$ lattice CG. Two sequences of two-qubit gate additions for $6$- and $8$-qubit IGs are utilized, and are shown in \cref{fig6}. We sequentially add cycle-like edges to IGs, starting from their path-like counterparts. Our aim is twofold: Firstly, we wish to characterize and associate the performance of our HQAA with the \textit{number of cycle-like edges} in the corresponding IG; furthermore, we are interested in the specific topological properties of said IGs. In this section, we added cycle-like edges in two different patterns: following a \textit{depth-first} pattern, which seeks to saturate the vertex degree of one vertex of the IG before adding more edges to the next, and \textit{breadth-first} pattern, in which edges are added in such a way that the vertex degrees of all nodes stay approximately equal.

In the small-$n$ regime (for an $n \times n$ QPU CG), BFAAs can be utilized, as the size of the coupling and IGs are sufficiently small. Even so, it must be stated that in all of our simulation results, our BFAA requires several orders of magnitude more time in order to complete each trial than the HQAA or TAA. 

Cycle-like chords that were sequentially added to $s$-qubit path-like IGs are shown in \cref{fig6}; here, $s \in \{4,6,8\}$. Due to the fact that there are several different ways in which one may add chords to the six- and eight-qubit quantum algorithms, two different ways of edge addition were used for the respective $s$-qubit cycles, in order to explore and compare two sets of IGs that are equivalent in all ways, except when viewed through the lens of \textit{topological graph equivalence}. We name these resulting sets of IGs generated by each iterative procedure as \textit{breadth-first} or \textit{depth-first} sequences. As shown in Sequence I (\cref{fig6}), a cycle is immediately created upon adding an edge between the first and last vertices of a path-like IG; subsequently, chords are added to the $s$-qubit cycle such that the degree of every sequentially-numbered vertex is maximized before proceeding to add chords to the consecutively-numbered vertices. We refer to this style of adding edges as a \textit{depth-first} approach. Sequence II, however, exhibits a \textit{breadth-first} approach to cycle-like edge addition, as chords are added in a manner such that the vertex degree of all vertices remain approximately equal. The red-highlighted edges represent newly-added chords to a particular sequence.

One may ask what the reasoning behind such an elaborate edge-generation procedure may be; after all, is it not sufficient to only consider the numbers of virtual qubits and gate overhead for understanding the qubit-assignment properties of a quantum algorithm using an HQAA? One of the main purposes of this manuscript is to show that topological-graph properties such as \textit{degree centrality} play an important role in qubit-assignment when using an HQAA; as such, it was imperative to construct sets of interaction-graph benchmarks which exhibit the same numbers of two- and single-qubit gates, exhibiting their only differences in their topological-graph properties. As an example, consider a map $f$ from graph d6) in Sequence I (which for this example we shall denote $X$ for simplicity) to graph d6) in Squence II (we here we shall denote $Y$ for simplicity). According to the definition of \textit{topological equivalence} introduced in \cref{background-qmapping-section}, three main conditions must be fulfilled, in order to define such a map $f:X \mapsto Y$. It should be obvious that condition 1 does not hold, as their graph Laplacians \cite{wilson1} do not share the same eigenvalue spectrum\footnote[4]{It is well-known in spectral graph theory \cite{KnauerKnauer+2019,wilson1,MERRIS1994143}) that two graphs $X$ and $Y$ are isomorphic (i.e. exhibit an edge-set and vertex-set bijection) if a permutation matrix $\mathcal{P}$ exists for which $\mathcal{P}X\mathcal{P}^{-1} = Y$; a necessary consequence of this is that both graph Laplacians share the same eigenvalue spectrum. Since we know a priori that $X$ and $Y$ do not share the same eigenvalue spectrum, we can conclude that no permutation matrix exists such that a vertex- and edge-set bijection exists; thus condition 1 does not hold.}. In our simulations, we hold the total numbers of single- and two-qubit gates to be the same for our benchmarks, but allow for different topological structures to emerge between IGs, as shown by \cref{fig6}. In this way, we have positioned ourselves such that we can now evaluate the performance of a HQAA relative to varying the topological-graph properties of the IGs, while keeping the coupling-graph structure constant.

\begin{figure*}
\centering
\includegraphics[width=\textwidth]{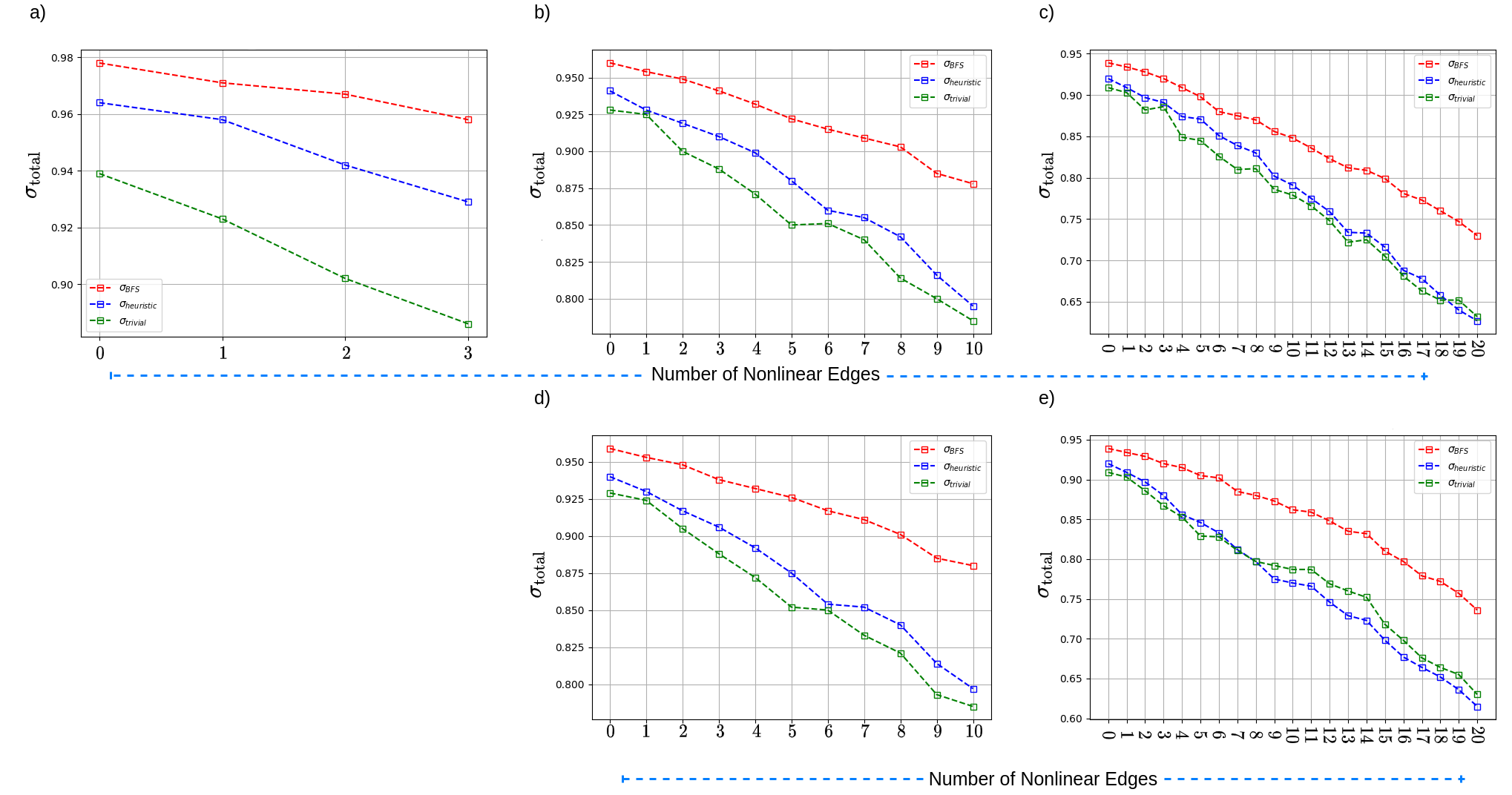}
\caption{a)-c) show the results for the cycle-like benchmarks in \cref{fig6} for Sequence I (the depth-first benchmarks); the x-axis of the figure follows the sequence of cycle-like edge additions. The average runtime per trial for all of the benchmarks was approximately $60$ seconds, $0.5$, and $0.2$ milliseconds for the BFAA, HQAA, and TAA, respectively. The total runtime for the entire simulation was approximately 60 hours. d)-e) show the results for the cycle-like benchmark IGs in \cref{fig6} for Sequence II. These breadth-first benchmarks exhibited an average runtime per trial were approximately equal to those stated for the depth-first benchmarks.}
\label{fig7}
\end{figure*}

The results for $4$-, $6$-, and $8$-qubit IGs that were assigned in accordance with the IGs in Sequence I and II are depicted in \cref{fig7}. \cref{fig7}a, \cref{fig7}b and \cref{fig7}c represent the results procured from IGs in Sequence I; \cref{fig7}d and \cref{fig7}e represent the Sequence II results.  In each subfigure, the BFAA, HQAA, and TAA results are shown in red, blue, and green, respectively. The success rates of each assignment algorithm $\{\sigma_{\text{brute}},\sigma_{\text{heuristic}},\sigma_{\text{trivial}}\}$ are graphed as a function of the number of cycle-like edges that have been added from the original sequence start in Ia4), Ia6), and Ia8) in \cref{fig6}. The average runtime per trial for the BFAA, HQAA, and TAA are on the order of about $60$ seconds for the BFAA versus $0.5$ and $0.2$ milliseconds for the HQAA and TAA, respectively. These figures largely stay the same for the Sequence II benchmarks, as the BFAA exhibits an average solution time of several orders of magnitude higher than the other two qubit-assignment algorithms. 

\cref{fig7}a shows that our HQAA's success rate can approach the BFAA's for $4$-qubit IGs, no matter how many cyclic chords are added; indeed, it is observed that even a quantum algorithm representing a $K_{4}$ IG can be effectively assigned to the $3 \times 3$ lattice CG in our simulations. In contrast, \cref{fig7}b and \cref{fig7}c depict performance decreases for our heuristic optimization as more and more cycle-like edges are added to the IG; \cref{fig7}c, however, is unique in that the performance of the HQAA achieves roughly analogous performance to that of the TAA, until finally, after $18$ cycle-like edges have been added, a "critical point" is reached. We define this critical point to be the point at which the HQAA's success rate matches that of the TAA's. This behavior is not witnessed in the $6$-qubit IGs; indeed, although the $6$-qubit benchmarks do exhibit steadily-decaying performance, our results indicate that this tendency is more prevalent with larger quantum circuits, such as the $8$-qubit benchmarks. 

The breadth-first benchmarks of Sequence II exhibit a distinct trend from the first-sequence IGs. Although \cref{fig7}d largely appears to agree with \cref{fig7}b, the second-sequence IGs in \cref{fig7}e exhibit much worse HQAA success rates, compared to the first-sequence alternatives. In point of fact, the critical point appears sooner in our simulations while following Sequence II. Approximately $18$ cycle-like edges were added in the depth-first analysis of Sequence I; in contrast, the point wherein the HQAA's success rate is equal to that of the TAA's is observed after approximately $7$-$8$ cycle-like edges have been added from Sequence II (breadth-first).

A few comments are in order here. Firstly, the data obtained from these simulations reveal that our HQAA can adequately assign IGs which exhibit a low degree of cycle-like edges, provided that the quantum circuit in question is not very deep. As we increase the sheer amount of the two-qubit gate invocations between interaction-graph qubits, it can be seen that our HQAA can tolerate a small amount of interaction-graph cyclicity, especially if this cyclicity is centralized on a subset of the total vertices. Next, and perhaps more importantly, for a given number of cycle-like edges added to a path-like IG, the success rate varies significantly for our HQAA. However, if we take two IGs from Sequences I and II with equivalent numbers of edges, we found that the local distribution of edge assignments play a role in the success rates observed, as these localized edge-assignment discrepancies are the only difference between any two pairs of IGs from Sequence I and II. This local distribution discrepancy generates the observed topological inequivalence \textit{globally} for the IGs, and plays a significant role in the performance of the HQAA.

These same simulations were performed for the same benchmarks, with the only difference being that the number of total gates for each benchmark was multiplied by two and four. Our results largely conform with those shown in \cref{fig7}, as only the scaling of the success-rate metric changes. This observation is indeed expected, as the design of the experiment and the cost function themselves facilitate such a rescaling.

Lastly, as shown in \cref{fig7}c and \cref{fig7}e, the heuristic's and TAA's success rates are relatively close. One may surmise that a reason for this proximity is related to the high occupancy of the coupling-graph qubits during our simulations ($8/9$ of the available QPU qubits were utilized). The next section will provide information related to the large path-like interactions graphs that were tested and how they scale in the regime $n > 3$ for a corresponding $n \times n$ QPU CG. We specifically address the question of how the occupancy percentage of the QPU affects our HQAA's solution.

\subsection{Scaling Properties of our HQAA on larger QPU Lattices}\label{large-linear-section}

\begin{figure*}
\centering
\includegraphics[width=\textwidth]{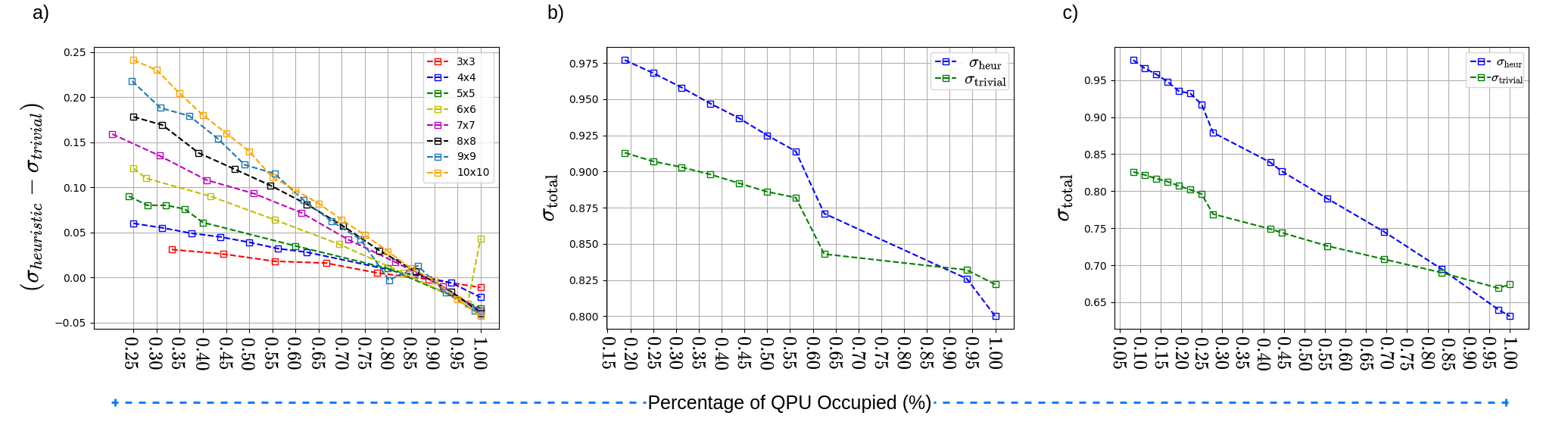}
\caption{Success-rate differences between the success rates of the HQAA and TAA $(\sigma_{\text{heurstic}}, \sigma_{\text{trivial}})$ taken from $n$-qubit path-like IGs which have been assigned to $n \times n$ QPU CGs. a) displays the success rates differences without any noise scaling. b) and c) detail the success rate $\sigma_{\text{total}}$ calculated for both $n = 4$ and $n = 6$ QPU CGs, respectively.}
\label{fig8}
\end{figure*}

The benchmarks tested in this section are all path-like, but vary in size so as to occupy different percentages of an $n \times n$ QPU CG, where $n = 3,\dots,10$. A CG of lattice dimensions $n \times n$ is initialized, and sequentially larger and larger path-like IGs are assigned to a CG until it is $100\%$ occupied; as in the previous simulations, vertices of the CG are taken to be qubits. Addtionally, no explicit noise scaling was utilized as the CG increases; the magnitudes of noise for two-qubit and single-qubit error rates, as well as measurements are kept the same as described in \cref{description_algs-section}. As BFAAs were not utilized for this section (becoming practically intractable if $n > 3$), success rates were instead compared between the HQAA and TAA, as shown in \cref{fig8}b and \cref{fig8}c.

The results from our study are displayed in \cref{fig8}. \cref{fig8}a shows the success-rate difference $\{\sigma_{\text{heuristic}}-\sigma_{\text{trivial}}\}$, calculated between our HQAA and the TAA, with respect to path-like IGs that progressively fill the entire QPU CG. The different colors in \cref{fig8}a denote differing $n$-values for the lattice dimensions of the CG. \cref{fig8}b and \cref{fig8}c display the actual success rates calculated under no noise scaling behavior, for $4 \times 4$ and $6 \times 6$ coupling-graph dimensions, respectively.

In these graphs, one can observe several trends: firstly, as the $n$-value of the CG increases, the success-rate difference becomes steadily larger for path-like IGs that occupy the same percentage of the CG when fully assigned, until about $75\%$ of the QPU is occupied. Next, after approximately $75\%$ of the CG has been filled, the success-rate difference becomes negligible; this success-rate difference implies that an effectively negligible difference between the heuristic- and the TAA is observed. Lastly, after roughly $85-90\%$ of the lattice CG is occupied, the success-rate difference is not only negligible, but starts to dip below zero, the main observation here being that the HQAA cannot find a solution that outperforms the success rate measured from the TAA. We will discuss the consequences of this observation in \cref{discussion-section}.

Additionally, we performed the same tests for CGs with increasing $n$-values such that a uniform increase in noise of order $4$ times larger than the noise in \cref{fig8}, as well as under exponentially increasing noise. The purpose of these rescaled-noise simulations was to investigate potentially more realistic noise, as device error rates are expected to increase due to crosstalk as quantum processors become larger \cite{Hangleiter2020crosstalkdiagnosis,Murali_2020software}. Our results from these simulations largely confirm the same trends that were mentioned in the preceding paragraphs, albeit with steeper linear decays.

\section{Discussion}\label{discussion-section}

\begin{figure}[htp]
\centering
\hskip-0.5cm
\includegraphics[width=\columnwidth]{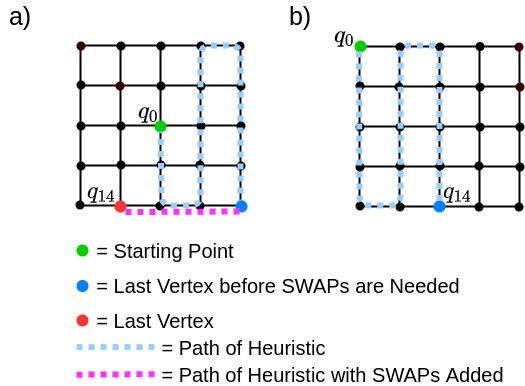}
\caption{An example scenario in which our HQAA could perform worse in comparison to a TAA. On the $5 \times 5$ CG pictured above, a) and b) represent possible final-assignment solutions for the HQAA and TAA, respectively. In a), one may initiate the assignment process in a highly-connected region of the CG; however, as one proceeds using the algorithm to consecutively assign the $15$ qubits in our quantum-circuit example, one may encounter a situation in which the heuristic-based solution may involve several more SWAP gates than normally anticipated. This situation is shown as a dotted line in blue (tracing out the nearest-neighbor assignment path of our HQAA) which essentially runs into a corner in the lower-right portion of the QPU CG, stopping at the vertex shown in blue. From this point forward, the next shortest-path distance would not be a nearest-neighbor vertex, and SWAPs will undoubtedly be needed in order to realize such a assignment solution (shown by the dotted line in magenta, which terminates at the red-labelled vertex $q_{14}$). In b), a TAA solution would better utilize the space and connectivity available for a QPU when less choices are available.}
\label{fig9}
\end{figure}

In this article, we have introduced an HQAA for the purpose of exploring the advantages and limitations for assigning topologically-inequivalent IGs to quantum hardware, in addition to systematically investigating the scaling behavior of our HQAA as the QPU size increases. The HQAA itself is \textit{noise-aware}, and the success rate is high, relative to a BFAA and TAA, which serve to effectively bound the performance of our HQAA from above and below. For small, low-depth quantum circuits, the HQAA is shown to provide a significant performance gain over a TAA solution, and in some cases for realistic benchmarks even approaches that of a BFAA. 

Two main stages of our work have led us to novel results about the behavior of simple HQAAs: Firstly, we have investigated the performance of our BFAA, HQAA, and TAA with low-depth quantum circuits which admit cycle-like IGs. Two particular edge-addition sequences were considered, both of which highlight the inherent limitations of the HQAA that we devised, albeit in different ways. The topological-graph properties of the cycle-like edges added to the circuit (i.e. depth-first or breadth-first edge assignments), as well as the sheer amount of cycle-like edges utilized, both were found to play a role in our HQAA's calculated success rate. These two observations are evidenced from an analysis of \cref{fig7}c and \cref{fig7}e: although our HQAA performs better than the TAA solution for most of the depth-first edge additions (Sequence I), we see that the performance is \textit{not much better}, implying that our HQAA can tolerate a relatively low amount of interaction-graph cyclicity, in the case of the number of qubits in the IG being relatively high. This observation comes as no surprise, seeing as the algorithm was designed to accommodate path-like IGs. Furthermore, for two IGs with the same number of cycle-like edges, it was found that our HQAA's performance depends on the particular manner in which the cycle-like interaction-graph edges are distributed. We verified that this fact is reinforced for deeper algorithms by running simulations with larger-depth IGs of the same form as described in \cref{fig6}. More specifically, Sequence I and Sequence II differ from each other due to their \textit{vertex centrality}. When looked at from this perspective, it is unsurprising that our HQAA, tailored to take into account vertex centrality, performs better for IGs that additionally exhibit . What is in fact novel about these results is \textit{how quickly} the performance of our HQAA falls off in comparison to another, topologically-inequivalent IG, and also in comparison to our approximate upper- and lower-bounds, provided by the BFAA and TAA. This falloff demonstrates that a topological-graph dependency is exhibited for our particular assignment strategy employed.

Secondly, we investigated the scaling behavior of our HQAA in the regime $n > 3$ for $n \times n$ QPU CGs. In this regime, our BFAA solutions to the assignment problem become intractable; as such, the HQAA was compared to the TAA described in \cref{description_algs-section}. Our novel results indicate that the HQAA scales well for quantum circuits with path-like IGs, as long as less than approximately $75\%$ of the QPU CG is filled (in comparison to our TAA). If one occupies more of the available space on the QPU, one can expect a trivial benefit from utilizing our HQAA, until approximately $85\%$ of the processor is allocated; after this marker, performance losses can be expected (with respect to our TAA), as our results denoted. Additionally, for comparable percentages of the CG that are filled, one can expect higher success rates relative to the TAA solution as $n$ is steadily increased. These same simulations were additionally employed for uniformly and exponentially increasing noise parameters, concomitant with observations that larger QPU devices experience more problems with error rates due to crosstalk \cite{Hangleiter2020crosstalkdiagnosis,Murali_2020software}. Our results confirm and reinforce the remarks stated above. 

One may ask why our HQAA tends to underperform as we occupy larger percentages of a QPU CG. An answer to this question can be seen when one considers that, as the CG is filled up, less and less nearest-neighbor choices are left for the HQAA to evaluate. As the algorithm itself functions by selecting first the maximum-degree vertex of the minimal error-rate edge, the heuristic-based solution will necessitate more SWAP gates as the processor is further allocated. In this regime (i.e.\ when over $75\%$ of the QPU is occupied), the TAA used in this study would be expected to outperform our HQAA for quantum circuits that give rise to path-like IGs. In this sense, using notions of vertex centrality for an HQAA's strategy may lead to detrimental results in the limit when over $75\%$ of the processor is filled. An example of this difficulty can be seen in \cref{fig9}; indeed, in a) and b), one sees the result of a greedy shortest-path assignment applied to a $5 \times 5$ CG. In \cref{fig9}a, we may start with a high-degree vertex (shown in green); as the algorithm proceeds to find nearest-neighbor solutions, the algorithm's attempt to assign the $15$-qubit path-like IG in our example may run into a portion of the device that is less-highly connected (shown by the blue-dotted line which terminates at the blue-labelled vertex). When such an event happens, the HQAA will continue searching for the nearest possible neighbor that is free; unfortunately, in this case several SWAP gates would be needed in order to realize the mapping shown (denoted by the magenta-dotted line that terminates at the red vertex $q_{14}$). In \cref{fig9}b, we see an example of our TAA would make better use of the space requirements for the $15$-qubit quantum circuit. One solution to this issue with our HQAA may be to utilize \textit{look-ahead} or \textit{look-behind} techniques, which would serve to match not only the nearest neighbor on the CG, but to additionally analyze several interaction-graph vertices \cite{palerwille,will-lookahead,2018tackling}. Such a qubit-mapping algorithm may improve overall success rates. In any case, it is clear that the choice of HQAA may or may not perform well for a given CG and cost function; we leave further discussion to future work.

As a final comment, it was observed that our BFAA scaled badly after $n = 3$ for the CGs tested; this does not necessarily imply that \textit{any exact simulation} is intractable. Although we note that finding an exact solution to the qubit-mapping problem is NP-hard, one may be able to write an exact mapping algorithm which does in fact scale better than ours for larger $n$-values.

Various other qubit-mapping algorithm proposals exist as well \cite{lao-mapping1,2021spectralgraphtheory,temporal1,hardware-aware,qubitproblem1,aliroquantum,ML1,ML2,bandic}; many of these could be analyzed and classified, as it is imperative to better understand which types of qubit-assignment algorithms may be most amenable for certain structural classes of quantum algorithms. In the interim, we expect that most (if not all) HQAAs will exhibit a topological-graph dependence via the IGs utilized, although this must be explored and verified in future work. Additionally, profiling the topological-graph properties of further benchmark sets such as those from \cite{revlib,queko} may serve to inform future proposals for QAAs, as well as more general qubit-mapping approaches. In this way, understanding the topological-graph properties of the qubit-mapping problem may serve to elucidate a more fundamental mathematical metric to quantify the performance of qubit-mapping strategies.

\section{Acknowledgements}

We thank Jens Eisert for useful discussions. MS would like to thank HQS Quantum Simulations GmbH (where a portion of this work was completed), and Prakesh Murali for providing the realistic benchmark files from \cite{noisy1} which were utilized in \cref{real-benchmarks-section}.

\clearpage

\bibliographystyle{unsrt}
\bibliography{bibliography}

\EOD

\end{document}